\documentclass{article}

\usepackage[preprint]{spconf}
\usepackage{amsmath,graphicx}

\usepackage{multirow}
\usepackage{amssymb}
\usepackage[bookmarks=false]{hyperref}
\usepackage{textcomp}

\copyrightnotice{\copyright\ 2023 IEEE}
\toappear{To appear in {\it Proc.\ ICASSP 2023, June 4-10, 2023, Rhodes Island, Greece}}

\title{On Negative Sampling for Contrastive Audio-Text Retrieval \vspace{-6pt}}
\name{Huang Xie, Okko R\"as\"anen, Tuomas Virtanen \vspace{-6pt}}
\address{Unit of Computing Sciences, Tampere University, Finland \vspace{-18pt}}
\begin{document}
%\ninept
    \setlength{\abovedisplayskip}{3pt}
    \setlength{\belowdisplayskip}{3pt}
    \maketitle
    \begin{abstract}
        This paper investigates negative sampling for contrastive learning in the context of audio-text retrieval.
        The strategy for negative sampling refers to selecting negatives (either audio clips or textual descriptions) from a pool of candidates for a positive audio-text pair.
        We explore sampling strategies via model-estimated within-modality and cross-modality relevance scores for audio and text samples.
        With a constant training setting on the retrieval system from~\cite{Xie2022Unsupervised}, we study eight sampling strategies, including hard and semi-hard negative sampling.
        Experimental results show that retrieval performance varies dramatically among different strategies.
        Particularly, by selecting semi-hard negatives with cross-modality scores, the retrieval system gains improved performance in both text-to-audio and audio-to-text retrieval.
        Besides, we show that feature collapse occurs while sampling hard negatives with cross-modality scores.
    \end{abstract}
    \vspace{-10pt}

    \section{Introduction}\label{sec:introduction}
    \vspace{-6pt}

    Audio-text retrieval refers to retrieving audio or descriptive text that is relevant to a given query from the other modality.
    It has great potential in real-world applications, such as search engines and multimedia databases.
    Early works~\cite{Slaney2002Semantic, Chechik2008Large, Ikawa2018Acoustic, Elizalde2019Cross} have focused on audio retrieval with separate words, for example, using words ``horse trot'' to search for audio containing sounds like horse trotting or galloping.
    Real-world audio inherently consists of sounds distributed across the temporal axis.
    Retrieving audio with separate words usually emphasizes on the presence of certain sounds and neglects their temporal information (e.g., relative positions on the temporal axis).
    A more natural way for humans to describe the desired data is by natural language descriptions.
    For example, a detailed description ``a woman talks followed by a dog barking'' provides more information than words like ``human voice'' and ``dog barking''.
    This paper concentrates on audio-text retrieval with unconstrained textual descriptions.

    With the availability of audio-caption datasets~\cite{Drossos2020Clotho, Kim2019AudioCaps}, several works have explored audio-text retrieval with free-form textual descriptions (i.e., audio captions).
    \mbox{Oncescu~\textit{et al.}~\cite{Oncescu2021Audio}} first established benchmarks in this topic by adapting video retrieval models.
    Our previous work~\cite{Xie2022Unsupervised} investigated audio-text retrieval by learning alignment between audio and their corresponding captions.
    \mbox{Mei~\textit{et al.}~\cite{Mei2022On}} evaluated the popular learning objectives (e.g., triplet losses~\cite{Bromley1993Signature, Schroff2015FaceNet}) for training audio-text retrieval models.
    Recently,~\mbox{Deshmukh~\textit{et al.}~\cite{Deshmukh2022Audio}} proposed a contrastive learning framework for audio-text retrieval, where they combined the latest advancements in audio models (e.g., CNN14~\cite{Kong2020PANNs}, HTSAT~\cite{Chen2022HTSAT}) with a pretraining approach named CLAP~\cite{Elizalde2022CLAP}.
    They also introduced a new collection of audio-text pairs, which was referred to as WavText5K\@.
    Besides, the newly introduced~\textit{language-based audio retrieval} task~\cite{Xie2022Language} in DCASE 2022 Challenge received a total number of 31 submissions, which showed an increasing interest in this topic from the audio research community.

    Most of the literature tackle audio-text retrieval with contrastive learning methods, i.e.,~\textit{contrastive audio-text retrieval}.
    A dual-encoder architecture, consisting of an audio encoder and a text encoder, has been widely employed~\cite{Xie2022Language}.
    Audio data and textual descriptions are encoded as embeddings into a common multimodal space.
    By optimizing a contrastive learning objective (e.g., triplet loss), the relevant audio and text embeddings are pulled close to each other, while the irrelevant ones are pushed far away from each other in the shared space.
    The aforementioned works~\cite{Oncescu2021Audio, Xie2022Unsupervised, Mei2022On, Deshmukh2022Audio} have mainly focused on the aspects such as retrieval architecture, learning objective, pretraining approach, and audio-text data collection.
    In contrast, the selection of negative samples for contrastive learning, i.e., negative sampling (NS), is under-explored.
    Previous works~\cite{Schroff2015FaceNet, Robinson2021Contrastive} from computer vision show that the choice of negative samples has a decisive impact on the success of contrastive learning.
    For example, most negative samples are easy to discriminate, having minor contributions to model training~\cite{Robinson2021Contrastive}, and some are even counterproductive, leading to a collapsed model~\cite{Schroff2015FaceNet}.

    This work aims to study and compare different strategies for negative sampling in contrastive audio-text retrieval.
    Particularly, we explore negative sampling via model-estimated within-modality and cross-modality relevance scores between audio and text samples in a mini-batch.
    With a constant training setting on the retrieval system from~\cite{Xie2022Unsupervised}, we study eight sampling strategies, including hard and semi-hard negative sampling.
    Experimental results show that retrieval performance varies dramatically among these strategies.
    We empirically demonstrate that employing an appropriate strategy for negative sampling (e.g., selecting semi-hard negatives with cross-modality scores) can result in improved performance.
    \vspace{-10pt}

    \section{Contrastive Audio-Text Retrieval}\label{sec:contrastive-audio-text-retrieval}
    \vspace{-6pt}

    \begin{figure*}[!t]
        \centering
        \includegraphics[width=1.0\textwidth]{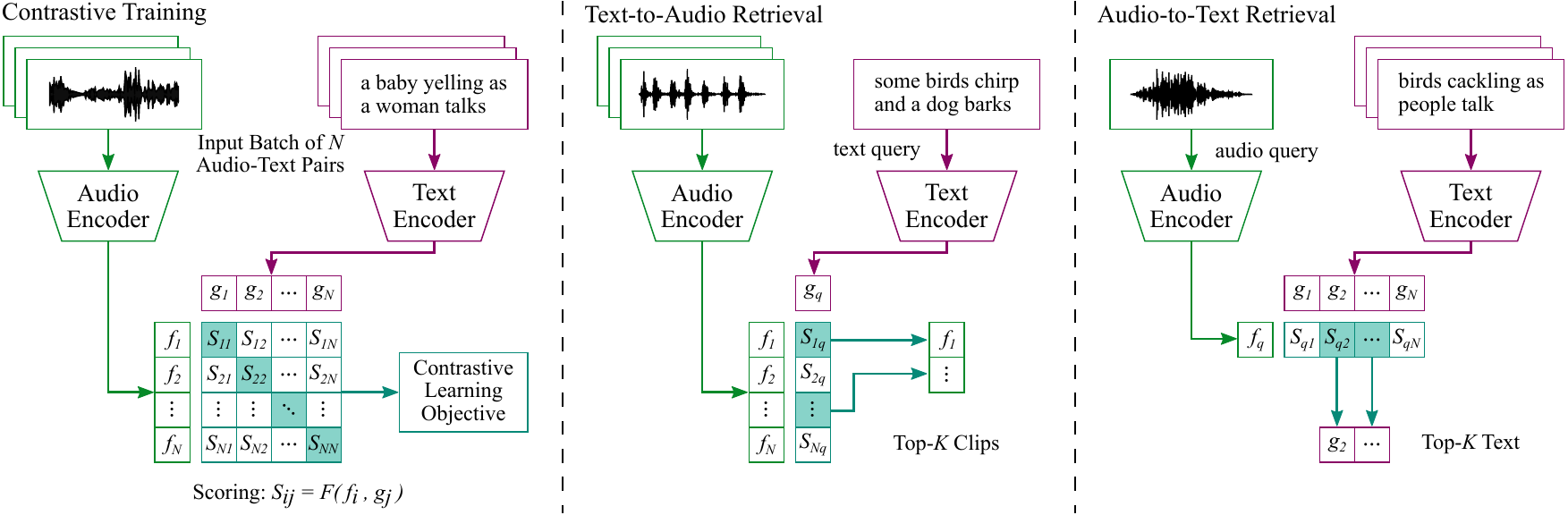}
        \vspace{-18pt}
        \caption{Contrastive audio-text retrieval framework. From left to right: 1) Contrastive training. The dual-encoder framework is trained on a batch of $N$ audio-text pairs by optimizing a contrastive learning objective (e.g., triplet loss). 2) Text-to-audio retrieval. For a given text query (e.g., ``some birds chirp and a dog barks''), the top-$K$ audio clips that have a high score $S$ with the text query are retrieved. 3) Audio-to-text retrieval. For a given audio clip, the top-$K$ textual descriptions that have a high score $S$ with the audio query are retrieved.}
        \label{fig:contrastive_retrieval_framework}
        \vspace{-10pt}
    \end{figure*}

    In this section, we formulate the task of contrastive audio-text retrieval with a model-agnostic dual-encoder framework, as illustrated in Figure~\ref{fig:contrastive_retrieval_framework}.
    Let $X=\{(x_{i},y_{i})\}_{i=1}^{N}$ be a batch of $N$ audio-text pairs, where the $i$-th audio clip $x_{i}$ is described with the $i$-th textual description $y_{i}$.
    Due to the lack of graded audio-text relevance in existing datasets (e.g., Clotho~\cite{Drossos2020Clotho}, AudioCaps~\cite{Kim2019AudioCaps}), we consider only binary relevance between audio clips and textual descriptions.
    Thus, $(x_{i},y_{i})$ is regarded as a positive pair and $(x_{i},y_{j})$ with $i \neq j$ as a negative pair.

    The audio-text pairs are projected into a common representation space via the dual-encoder framework.
    Let $\theta(\cdot)$ be the audio encoder, and $\phi(\cdot)$ be the text encoder.
    With the projected representations $f_{i}=\theta(x_{i})$ and $g_{j}=\phi(y_{j})$, a function $F(\cdot)$, written as
    \begin{equation}
        \label{eq:relevance_function}
        S_{ij} = F(f_{i},g_{j}),
    \end{equation}
    is defined to measure the semantic relevance between $x_{i}$ and $y_{j}$ with a score $S_{ij}$.
    The popular choices of $F(\cdot)$ include dot product~\cite{Xie2022Unsupervised, Deshmukh2022Audio} and cosine similarity~\cite{Oncescu2021Audio, Mei2022On}.

    The encoders $\theta(\cdot)$ and $\phi(\cdot)$ are trained on $X$ with a contrastive learning objective $\mathcal{L}$.
    By optimizing $\mathcal{L}$, $f_{i}$ and $g_{i}$ are pulled close to each other (i.e., being of high relevance), while $f_{i}$ and $g_{j}$ with $i \neq j$ are pushed far away from each other (i.e., being of low relevance).
    % By optimizing $\mathcal{L}$, representations of a positive audio-text pair are pulled close to each other (i.e., being of high relevance), while those of a negative pair are pushed far away from each other (i.e., being of low relevance).
    The common choices of $\mathcal{L}$ include triplet loss~\cite{Xie2022Unsupervised, Oncescu2021Audio, Xie2022Language} and symmetric cross-entropy loss~\cite{Mei2022On, Deshmukh2022Audio}.
    The triplet loss takes as input triplets of samples, consisting of one negative sample along with one positive audio-text pair, while the symmetric cross-entropy loss includes multiple negative samples for every positive pair.
    To alleviate the influence of sample quantity on evaluation and investigate the effect of sample quality, we utilize the instance-based triplet loss~\cite{Xie2022Unsupervised} (see Section~\ref{subsec:retrieval-system}).
    For audio-text retrieval, audio or text samples that have a high score $S$ with the given query from the other modality are retrieved, as illustrated in Figure~\ref{fig:contrastive_retrieval_framework}.
    \vspace{-10pt}

    \section{Negative Sampling}\label{sec:negative-sampling}
    \vspace{-6pt}

    This section presents eight different strategies for negative sampling in contrastive audio-text retrieval.
    To avoid comparing positive pairs with all negative samples (i.e., heavy computational burden), we apply these strategies to select negative samples within a mini-batch of audio-text pairs, i.e., mini-batched negative sampling.
    These eight strategies are grouped into two categories: basic and score-based.
    \vspace{-12pt}

    \subsection{Basic Negative Sampling}\label{subsec:basic-ns}
    \vspace{-3pt}

    With sampling strategies in this category, each negative sample has the same chance of being selected.

    \textbf{Random Negative Sampling}.
    The simple random strategy works with a uniform distribution over negative samples in either modality, and selects one negative sample from each modality for every positive audio-text pair.
    It is the most basic strategy for negative sampling, and commonly used as the default setup in contrastive learning~\cite{Xu2022Negative}.
    Therefore, we employ this strategy as the baseline method in this work.

    \textbf{Full-Mini-Batch Negative Sampling}.
    The full-mini-batch strategy arbitrarily selects all negative samples within the same mini-batch for each positive pair.
    It is generally believed that contrastive learning benefits from increased sample size~\cite{Awasthi2022Do}.
    With this strategy, more negative samples contribute to training the dual-encoder framework.
    \vspace{-12pt}

    \subsection{Score-based Negative Sampling}\label{subsec:score-based-ns}
    \vspace{-3pt}

    Previous works~\cite{Schroff2015FaceNet, Robinson2021Contrastive} show that informative negative samples promote model optimization.
    The term~\textit{hardness} is commonly used to represent how informative a negative sample is for a positive pair.
    For example, negative samples that are difficult to distinguish from positive ones are often mentioned as~\textit{hard negatives}~\cite{Robinson2021Contrastive}, which are intuitively informative.

    We define the hardness of a negative sample by its score on a positive audio-text pair.
    Specifically, we compute within-modality and cross-modality scores for negative audio and text samples with~\eqref{eq:relevance_function}.
    For an audio-text pair $(x_{i},y_{i})$, a text sample $y_{j}$ has a within-modality score
    \begin{equation}
        \label{eq:text_relevance}
        S_{ij}^{text} = F(g_{i},g_{j}).
    \end{equation}
    Similarly, an audio sample $x_{k}$ has a within-modality score
    \begin{equation}
        \label{eq:audio_relevance}
        S_{ik}^{audio} = F(f_{i},f_{k}).
    \end{equation}
    Their respective cross-modality score $S_{ij}$ and $S_{ki}$ are calculated directly with~\eqref{eq:relevance_function}.

    \textbf{Within-Modality Hard Negative Sampling}.
    This strategy treats negative samples that have the highest within-modality score, along with their paired counterpart, as hard negatives.
    We experiment with selecting hard negatives based on either $S^{text}$ (i.e., text-based) or $S^{audio}$ (i.e., audio-based).
    For example, in text-based hard negative sampling, the text sample $y_{j}$ having the maximal $S_{ij}^{text}$, together with its paired audio $x_{j}$, are the hard negatives for $(x_{i},y_{i})$.
    For comparison, we also experiment with easy negatives, i.e., negative samples that have the lowest within-modality score.

    \textbf{Cross-Modality Hard Negative Sampling}.
    In this strategy, negative samples that have the highest cross-modality score on a positive pair are selected as hard negatives for experiment.
    For an audio-text pair $(x_{i},y_{i})$, we collect its hard negatives $y_{j}$ in text modality (i.e., having maximal $S_{ij}$) and $x_{k}$ in audio modality (i.e., having maximal $S_{ki}$).

    \textbf{Cross-Modality Semi-Hard Negative Sampling}.
    The term~\textit{semi-hard negative} was originally coined in face recognition~\cite{Schroff2015FaceNet}.
    Negative face images, which had a distance $d_{neg}$ (e.g., squared Euclidean distance) away from the anchor similar to $d_{pos}$ of the positive one (i.e., $d_{pos} < d_{neg} < d_{pos} + \varepsilon$ with a margin $\varepsilon$), were called semi-hard.
    We adapt this idea with cross-modality scores to select semi-hard negative samples for a positive audio-text pair.
    Specifically, we take negative samples that have a cross-modality score closest to that of the positive pair as semi-hard negatives.
    For example, an audio-text pair $(x_{i},y_{i})$ has two semi-hard negatives: a text sample $y_{j}$ with minimal $\left| S_{ij} - S_{ii} \right|$, and an audio sample $x_{k}$ with minimal $\left| S_{ki} - S_{ii} \right|$.
    \vspace{-10pt}

    \section{Experiments}\label{sec:experiments}
    \vspace{-6pt}

    This section introduces our experimental setup with the Clotho v2 dataset~\cite{Drossos2020Clotho} and the retrieval system from~\cite{Xie2022Unsupervised}.
    \vspace{-12pt}

    \subsection{Dataset}\label{subsec:dataset}
    \vspace{-3pt}

    The Clotho v2 dataset~\cite{Drossos2020Clotho} consists of 5,929 audio clips, with each clip having five human written captions (i.e., 29,645 in total), which makes it naturally suitable for audio-text retrieval.
    Each clip lasts for 15--30 seconds, and every caption contains 8--20 words.
    This dataset is divided into three splits: a development split with 3,839 clips and 19,195 captions, a validation split with 1,045 clips and 5,225 captions, and an evaluation split with 1,045 clips and 5,225 captions, respectively.
    All audio clips are sourced from the Freesound platform~\cite{Font2013Freesound}, and captions are crowd-sourced using a three-step framework~\cite{Drossos2020Clotho}.
    Due to the fact of data imbalance (i.e., having more captions than audio clips), retrieving captions becomes more difficult than retrieving audio clips in this dataset.
    %\vspace{-12pt}

    \subsection{Retrieval System}\label{subsec:retrieval-system}
    \vspace{-3pt}

    We perform audio-text retrieval with the aligning framework from~\cite{Xie2022Unsupervised}, where a convolutional recurrent neural network (CRNN)~\cite{Xu2019ACRNN} is employed as the audio encoder and a pretrained Word2Vec (Skip-gram model)~\cite{Word2Vec_online} as the text encoder.
    This system is simple, trainable with triplet loss, and having few external dependencies (e.g., pretrained audio experts, external data for pretraining), which makes it convenient for experiment.
    Audio-text relevance scores are computed with audio frame embeddings and word embeddings~\cite{Xie2022Unsupervised}.

    \textbf{Audio Encoder}.
    The CRNN encoder~\cite{Xu2019ACRNN} extracts frame-wise acoustic embeddings from audio clips of variable length.
    Audio clips are pre-processed using a Hanning window of 40 ms with a hop length of 20 ms.
    Then, 64-dimensional log mel-band energies are extracted and fed into the CRNN encoder.
    For each audio clip, a sequence of 300-dimensional frame embeddings are generated.

    \textbf{Text Encoder}.
    Following~\cite{Xie2022Unsupervised}, we utilize the same pretrained Word2Vec~\cite{Word2Vec_online} as the text encoder.
    It includes 300-dimensional word embeddings for roughly three million case-sensitive English words.
    We convert captions into sequences of word embeddings in a word-by-word manner.

    \textbf{Triplet Ranking Loss}.
    The retrieval system is optimized by minimizing an instance-based triplet ranking loss~\cite{Xie2022Unsupervised}
    \begin{equation}
        \label{eq:triplet_ranking_loss}
        \begin{split}
            \mathcal{L} = \dfrac{1}{N} \sum_{i=1}^{N} & [ \max (0, S_{ij} - S_{ii} + 1)      \\
            & + \max (0, S_{ki} - S_{ii} + 1) ],
        \end{split}
    \end{equation}
    where $j$ ($j \neq i$) indexes the sampled negative caption $y_{j}$ and $k$ ($k \neq i$) the sampled negative clip $x_{k}$ for a positive audio-text pair $(x_{i},y_{i})$.
    Audio-text relevance score $S$ is computed by averaging trimmed dot products of frame and word embeddings~\cite{Xie2022Unsupervised}.
    For full-mini-batch NS, $S_{ij}$ and $S_{ki}$ are averaged over negative captions and clips of $(x_{i},y_{i})$, respectively.

    \begin{table*}[!t]
        \centering
        \begin{tabular}{r|l|ccc|ccc}
            \hline
            \multirow{2}{*}{\bfseries Category} & \multirow{2}{*}{\bfseries Strategy} & \multicolumn{3}{c|}{\bfseries Text-to-Audio}     & \multicolumn{3}{c}{\bfseries Audio-to-Text}       \\
            \cline{3-8}
            &                             & \bfseries mAP  & \bfseries R@5  & \bfseries R@10 & \bfseries mAP  & \bfseries R@5  & \bfseries R@10 \\
            \hline
            \multirow{2}{*}{Basic}       & Random NS                   & 0.057          & 0.074          & 0.129          & 0.030          & 0.018          & 0.036          \\
            & Full-mini-batch NS          & 0.054          & 0.064          & 0.120          & 0.030          & 0.019          & 0.037          \\
            \hline
            \multirow{6}{*}{Score-based} & Cross-modality Semi-hard NS & \textbf{0.121} & \textbf{0.171} & \textbf{0.274} & \textbf{0.046} & \textbf{0.030} & \textbf{0.058}  \\
            & Cross-modality Hard NS      & 0.007          & 0.005          & 0.010          & 0.004          & 0.001          & 0.002          \\
            \cline{2-8}
            & Text-based NS (hard)        & 0.065          & 0.083          & 0.148          & 0.027          & 0.017          & 0.031          \\
            & Text-based NS (easy)        & 0.028          & 0.033          & 0.057          & 0.018          & 0.011          & 0.021          \\
            \cline{2-8}
            & Audio-based NS (hard)       & 0.034          & 0.037          & 0.072          & 0.030          & 0.018          & 0.035          \\
            & Audio-based NS (easy)       & 0.011          & 0.005          & 0.010          & 0.005          & 0.003          & 0.005          \\
            \hline
        \end{tabular}
        \caption{Experimental results for different negative sampling strategies with the retrieval system.}
        \label{tab:experimental_results}
        \vspace{-10pt}
    \end{table*}

    \textbf{Training Setup}.
    We train the retrieval system with mini-batches of 32 audio-text pairs in the development split for at most 120 epochs, and monitor the loss~\eqref{eq:triplet_ranking_loss} on the validation split during training.
    An Adam optimizer with an initial learning rate of $0.001$ is adopted to optimize the training process.
    The learning rate is reduced by a factor of ten once the validation loss does not improve for five epochs.
    Training is terminated by early stopping with a patience of ten epochs.

    The trained system is used for audio-text retrieval with the evaluation split.
    Retrieval is performed bidirectionally between audio and text modalities: text-to-audio retrieval (i.e., retrieving audio clips that are relevant to a text query), and audio-to-text retrieval (i.e., searching for text descriptions pertaining to a given audio clip).
    All captions and audio clips in the evaluation split are used for retrieval.
    \vspace{-12pt}

    \subsection{Evaluation Metrics}\label{subsec:evaluation-metrics}
    \vspace{-3pt}

    Retrieval performance is measured in terms of mean average precision (mAP) and recall at $k$ (R@\textit{k} with \textit{k} $\in \{5, 10\}$).
    The mAP is the mean of average precision (AP) over all queries, with AP being the average of precisions at positions where relevant items are in a retrieved rank list.
    High mAPs usually indicate that relevant items are top-ranked in the retrieval results.
    The R@\textit{k} is defined as the proportion of relevant items among the top \textit{k} retrieved results to all the relevant items contained in the evaluation data, and averaged across all queries~\cite{Xie2022Unsupervised, Xie2022Language}.
    The more relevant items within top \textit{k} retrieved results, the higher R@\textit{k} it is.
    Note that, the R@\textit{k} used in previous works~\cite{Oncescu2021Audio, Mei2022On, Deshmukh2022Audio} measures the percentage of test queries for which the correct result is among the top \textit{k} retrieved results~\cite{Mithun2018Learning}.
    \vspace{-10pt}

    \section{Results and Analysis}\label{sec:results-and-analysis}
    \vspace{-6pt}

    This section presents the results for different NS strategies with a constant training setting on the retrieval system.

    \textbf{Overview}.
    The mAP and R@\textit{k} with \textit{k} $\in \{5, 10\}$ scores obtained with different NS strategies are present in Table~\ref{tab:experimental_results}.
    The results show that different NS strategies have a dramatic impact on system performance.
    For example, in text-to-audio retrieval, ``Cross-modality Semi-hard NS'' (i.e., sampling semi-hard negatives with cross-modality score) achieves a mAP of $0.121$ and recall scores of $0.171$ (R@5) and $0.274$ (R@10), which are better than those from random NS (i.e., $0.057$ in mAP, $0.074$ in R@5, and $0.129$ in R@10).
    Particularly, ``Cross-modality Semi-hard NS'' achieves the best performance in both retrieval tasks.
    It obtains a mAP of $0.046$ and recall scores of $0.030$ (R@5) and $0.058$ (R@10) in audio-to-text retrieval.
    Besides, all NS strategies obtain higher mAP and recall scores in text-to-audio retrieval.
    Since there are more captions than audio clips in the evaluation split (5,225 captions vs.\ 1,045 clips), audio-to-text retrieval becomes more difficult than its counterpart.

    We notice that there is a gap between our results and those from previous works~\cite{Mei2022On, Deshmukh2022Audio, Xie2022Language}.
    Note that, the state-of-the-art results~\cite{Mei2022On, Deshmukh2022Audio, Xie2022Language} usually rely on heavy pretraining with external data (e.g., AudioCaps~\cite{Kim2019AudioCaps}), advanced retrieval architecture (e.g., CLAP~\cite{Elizalde2022CLAP}), and efficient learning objective (e.g., symmetric cross-entropy loss~\cite{Mei2022On}).
    We believe that our study can contribute to these techniques to increase their performance.

    \textbf{Basic NS}.
    The random NS and full-mini-batch NS obtain similar results in both retrieval tasks, with mAP / R@5 / R@10 around $0.054$ / $0.064$ / $0.120$ in text-to-audio retrieval and $0.030$ / $0.018$ / $0.036$ in audio-to-text retrieval.
    Theoretically, more negatives are involved in training the retrieval system with full-mini-batch NS, which would have an impact on the performance.
    A possible explanation for this could be due to the small batch size (i.e., 32) used for experiment.

    \textbf{Score-based NS}.
    In contrast to ``Cross-modality Semi-hard NS'', ``Cross-modality Hard NS'' (i.e., selecting hard negatives with cross-modality score) performs worse in both tasks.
    For example, in text-to-audio retrieval, it obtains a mAP of $0.007$ and recall scores of $0.005$ (R@5) and $0.010$ (R@10), even worse than those from random NS\@.
    The outputs from the audio encoder show that feature collapse occurs with ``Cross-modality Hard NS'' (i.e., outputting zero vectors as acoustic embeddings), which would be caused by the high variance of gradients with too hard negatives.
    The remaining four strategies select either hard or easy negatives with within-modality score (i.e., text-based and audio-based).
    The results indicate that selection of within-modality hard negatives is more efficient than the use of easy negatives.
    \vspace{-10pt}

    \section{Conclusion}\label{sec:conclusion}
    \vspace{-6pt}

    In this work, we evaluated eight different negative sampling strategies for contrastive audio-text retrieval with a constant training setting on the retrieval system from~\cite{Xie2022Unsupervised}.
    The experimental results show that retrieval performance varies dramatically among different strategies.
    Particularly, by selecting semi-hard negatives with model-estimated cross-modality scores, the system achieves improved performance in both text-to-audio and audio-to-text retrieval.
    We also notice that feature collapse occurs while sampling hard negatives with cross-modality scores.
    Additionally, we demonstrated that hard negatives selected with within-modality scores (i.e., text-based and audio-based) are more informative than those easy ones.
    As future work, we consider exploring non-binary audio-text relevance and pretraining for audio-text retrieval.
    \vspace{-10pt}

    \section{Acknowledgement}\label{sec:acknowledgement}
    \vspace{-6pt}

    The research leading to these results has received funding from Emil Aaltonen foundation funded project ``Using language to interpret unstructured data'' and Academy of Finland grant no. 314602.
    We would like to thank Khazar Khorrami for her useful discussions.

    \bibliographystyle{IEEEbib}
    \bibliography{ms}

\end{document}